\documentclass{article}
\usepackage[T1]{fontenc} 
\usepackage[utf8]{inputenc} 
\usepackage{ismir,amsmath,cite,url}
\usepackage{graphicx}
\usepackage{color}
\usepackage{caption}
\usepackage{subcaption}
\def\lbd{}	

\title{Learning to Generate Piano Music With Sustain Pedals}

\multauthor
{Joann Ching$^{1,2}$ \hspace{1cm} Yi-Hsuan Yang$^{1,3}$} {\\
$^1$ Research Center for IT Innovation, Academia Sinica\\
$^2$ Center for Music Technology, Georgia Institute of Technology, USA \\
$^3$ Yating Music Team, Taiwan AI Labs\\
{\tt\small jching9@gatech.edu, yang@citi.sinica.edu.tw}
}

\def\authorname{J. Ching and Y.H. Yang}

\begin{document}

\maketitle
\begin{abstract}
Recent years have witnessed a growing interest in research related to the detection of piano pedals from audio signals in the music information retrieval community \cite{detection2017, sustainPed, Liang_2019, TTtranscription}.
However, to our best knowledge, recent generative models for symbolic music have rarely taken piano pedals into account. 
In this work, we employ the transcription model proposed by Kong \emph{et al.} \cite{TTtranscription} to get pedal information from the audio recordings of piano performance in the AILabs1k7 dataset \cite{hsiao2021compound}, and then modify the Compound Word Transformer proposed by Hsiao \textit{et al.} \cite{hsiao2021compound} to build a Transformer decoder that generates pedal-related tokens along with other musical tokens. 
While the work is done by using inferred sustain pedal information as training data, the result shows hope for further improvement and the importance of the involvement of sustain pedal in tasks of piano performance generations. 
\end{abstract}
\section{Introduction}\label{sec:introduction}
Research on automatic music generation dates back to the 80s. With the trend in recent days, results of such models have improved vastly \cite{huang2018music,hsiao2021compound,emopia}. However, while the usage of sustain pedal is a common practice in piano performances nowadays, there has not been much work that includes such elements to recent deep learning-based generation models, possibly due to the scarcity of data. 
Piano performances are comprised of not only compositional features such as pitch and duration but also performance features. Given the relatively mature work of automatic music generation with compositional features, we want to investigate adding piano pedals to generative models of music. 

Such features are difficult for a model to learn because of its subjective nature being based on a performer's interpretation, especially in classical piano music. Within the broad idea of classical music, there are sub-genres such as Baroque, Classical, and Romantic, all possessing their style of playing, not to mention the usage of pedaling. 
For example, techniques of using the sustain pedal, especially in classical music, can be categorized into three types: \textit{Anticipatory}, \textit{Rhythmic}, and \textit{Legato}. 
In current existing symbolic datasets, MAESTRO\cite{hawthorne2018enabling} is the only one that contains 
piano pedal information collected from real performances. However, the usage difference between the three types of techniques might be too vague for a machine learning model to distinguish.

\begin{figure}
    \centering
    \begin{subfigure}[t]{\linewidth}
        \centering
        \includegraphics[width=70mm]{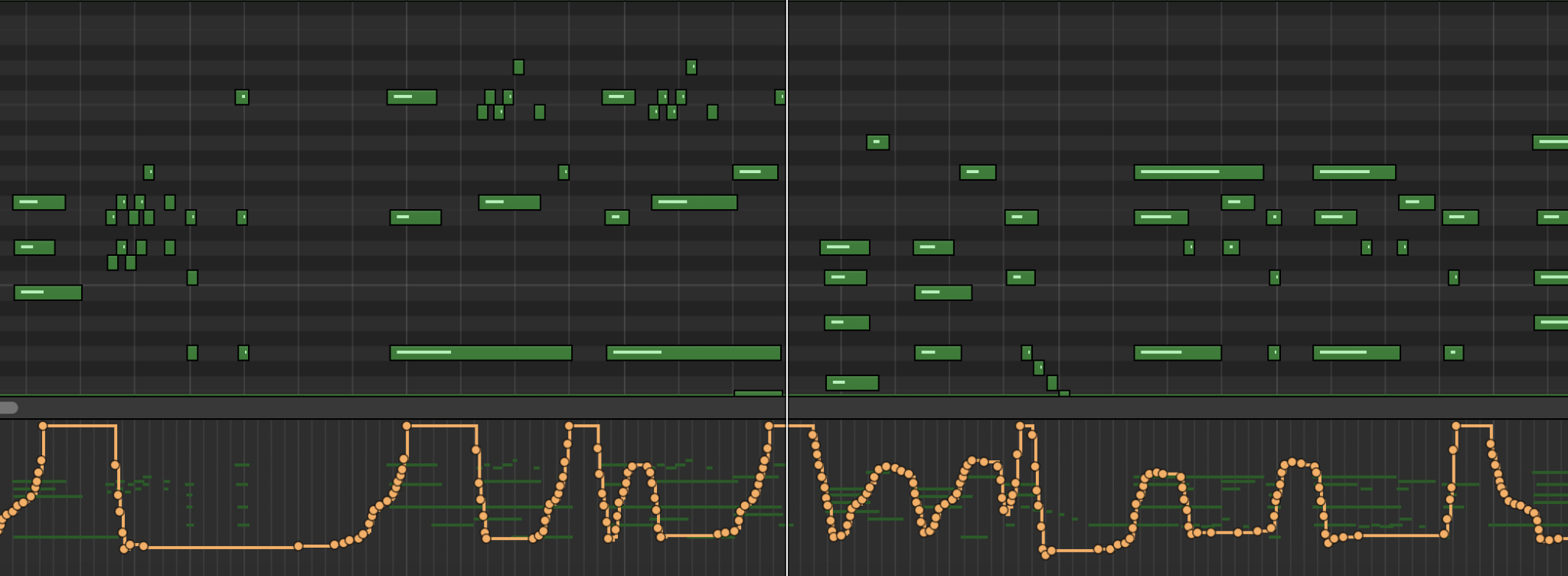}
        \caption{Visualization of a midi file from the MAESTRO dataset}
        \label{fig:classical}
    \end{subfigure}
    \hfill
    \begin{subfigure}[t]{\linewidth}
        \centering
        \includegraphics[width=70mm]{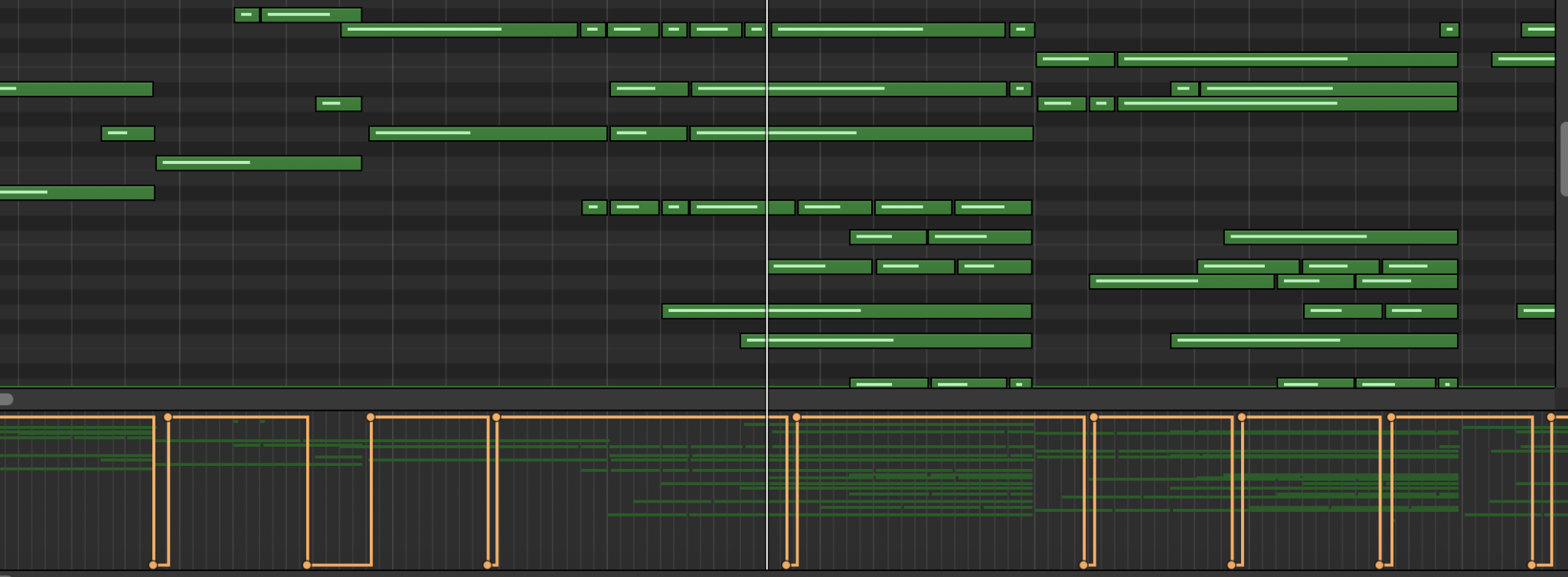}
        \caption{Visualization of a midi file from the AILabs1k7 dataset}
        \label{fig:pop}
    \end{subfigure}
    \caption{Illustration of the usage of the sustain pedal (orange lines) in MAESTRO \cite{hawthorne2018enabling} (classical; ground truth pedal labels) and AILabs1k7 \cite{hsiao2021compound} (pop, pseudo labels).}
    \label{fig:pedal}
\end{figure}

Although pedaling is arguably more important in classical music, we choose to start with using pop music as training data, 
given the simplicity of pedal techniques in pop music (e.g., pedal changes have a higher relationship with both beat and chord; see Fig. \ref{fig:pedal}). There are three types of pedals on a piano, the Soft Pedal, the Sostenuto Pedal, and the Sustain Pedal, but we only focus on the sustain pedal in this work. 
The source code for implementing the 
generation model can be found at a GitHub repo.\footnote{\url{https://github.com/joann8512/SusPedal-Gen}} Examples of generated pieces can be found at a demo webpage.\footnote{\url{https://joann8512.github.io/SusPedal-Gen/}}

\begin{figure}[t]
    \centering
    \begin{subfigure}{0.44\linewidth}
        \centering
        \includegraphics[width=35mm]{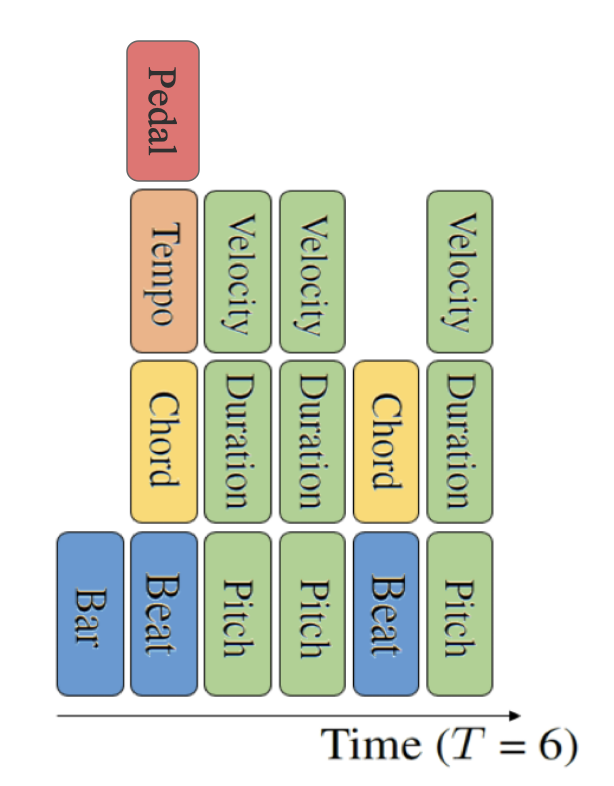}
        \caption{Grouped tokens}
        \label{fig:tokens}
    \end{subfigure}
    \hfill
    \begin{subfigure}{0.44\linewidth}
        \centering
        \includegraphics[width=35mm]{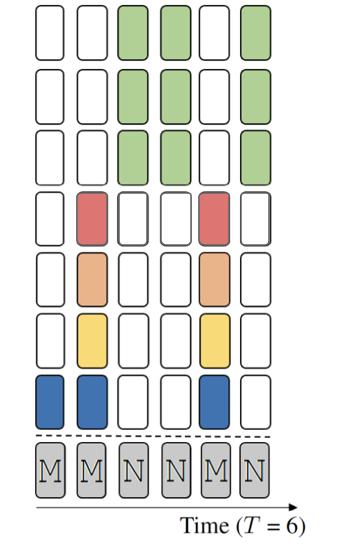}
        \caption{Compound word representation of grouped tokens}
        \label{fig:family}
    \end{subfigure}
    \caption{Example visualization of tokens}
    \label{fig:illustrate}
\end{figure}

\section{Dataset}
We use the AILabs1k7 dataset compiled by Hsiao \textit{et al.}\cite{hsiao2021compound} for training. The dataset contains 1,748 audio samples of pop piano cover music collected from the Internet, all of which are in 4/4 time signature (i.e., 4 quarter notes in each bar). We then transcribed the audio files into symbolic sequences using the open-source state-of-the-art high-resolution piano transcription model proposed by Kong \textit{et al.}\cite{TTtranscription} for its ability to transcribe sustain pedal information. 
It is possible to have undesirable cases from the transcription results for some of the songs, given that Kong \textit{et al.}'s model was trained for transcribing classical piano music. For example, notes are fragmented when the recording is engineered to have too many ambient effects (e.g., some songs in the EMOPIA dataset \cite{emopia}, according to our preliminary experiments). 
For AILabs1k7, the transcription result for pedals is acceptable.

\section{Data Representation}\label{sec:data_rep}

To train a Transformer model for generation,  we need a symbolic data representation that can be used as input to the model. The encoded symbolic data from the dataset are represented as ``event tokens'' to the model. Under the Compound Word Transformer, tokens with similar musical meaning are grouped to the same \textit{family}, which are then grouped into a \textit{super token} and placed on the same timestamp. The family groups contains \textit{metrical}, \textit{note}, and \textit{end of sequence}. 
In pop music, the most common and easiest way of pedaling is to change together with chord changes.  Therefore, we place the pedal tokens in the metrical family group, along with the chord and tempo tokens,  as depicted in Figure \ref{fig:illustrate}b. The presence of a pedal token signifies the onset time of using the sustain pedal.

The value taken by a pedal token indicates the pedal duration. 
We quantize the pedal duration and consider only the following 10 types of duration: 240, 480, 720, 960, 1440, 1920, 2400, 2880, 3360, 3840 ticks, where one beat corresponds to 480 ticks. The shortest is then an eighth note and the longest is 8 beats.
We apply quantization given that the transcribed pedal duration is not 100\% accurate anyway, and that reducing the vocabulary size of the pedal tokens (i.e., to 10) makes it easier for Transformers to learn the relations between notes and pedals.

\begin{table}[h]
\begin{tabular}{l|ll}
\hline
\textbf{}     & \textbf{Midi-like \cite{huang2018music}}       & \textbf{Ours} \\ \hline
Note onset    & \begin{tabular}[c]{@{}l@{}}Note-On\\ (0-127)\end{tabular}     & \begin{tabular}[c]{@{}l@{}}Note-On \\ (0-127)\end{tabular}    \\ \hline
Note offset   & \begin{tabular}[c]{@{}l@{}}Note-Off\\ (0-127)\end{tabular}    & \begin{tabular}[c]{@{}l@{}}Note-Duration\\ (1–64 32nd notes)\end{tabular} \\ \hline
Time grid     & \begin{tabular}[c]{@{}l@{}}Time-shift\\ (10–1000ms)\end{tabular}      & \begin{tabular}[c]{@{}l@{}}Subbeat (16 bins) \\ \& Bar (1 kind)\end{tabular}       \\ \hline
Tempo changes & ---       & \begin{tabular}[c]{@{}l@{}}Tempo\\ (56 kinds)\end{tabular}    \\ \hline
Chord         & ---       & \begin{tabular}[c]{@{}l@{}}Chord\\ (60 types)\end{tabular}    \\ \hline
Pedal         & \begin{tabular}[c]{@{}l@{}}Control Change\\ (2 kinds)\end{tabular} & \begin{tabular}[c]{@{}l@{}}Pedal\\ (10 types)\end{tabular}\\ \hline
\end{tabular}
\caption{Representation comparison on the ``MIDI-like'' event representation used by \cite{huang2018music} and our representation. Sizes of the token types are shown in brackets.}
\label{table:midi2token}
\end{table}

\section{Future Works}\label{sec:future}
In the current stage of the work, it is only proven that the results are acceptable. If the generation quality can be improved, with the sustain pedal information included in the generation process, the musical effects can be greatly improved in automatically generated music. For example, different emotions in music are often expressed through pedaling. In other words, pedaling in music generation could help to show the difference of emotions, and thereby creating more diversity in generated music.

\bibliography{ISMIRtemplate}

\begin{thebibliography}{1}
\providecommand{\url}[1]{#1}
\csname url@samestyle\endcsname
\providecommand{\newblock}{\relax}
\providecommand{\bibinfo}[2]{#2}
\providecommand{\BIBentrySTDinterwordspacing}{\spaceskip=0pt\relax}
\providecommand{\BIBentryALTinterwordstretchfactor}{4}
\providecommand{\BIBentryALTinterwordspacing}{\spaceskip=\fontdimen2\font plus
\BIBentryALTinterwordstretchfactor\fontdimen3\font minus
  \fontdimen4\font\relax}
\providecommand{\BIBforeignlanguage}[2]{{%
\expandafter\ifx\csname l@#1\endcsname\relax
\typeout{** WARNING: IEEEtran.bst: No hyphenation pattern has been}%
\typeout{** loaded for the language `#1'. Using the pattern for}%
\typeout{** the default language instead.}%
\else
\language=\csname l@#1\endcsname
\fi
#2}}
\providecommand{\BIBdecl}{\relax}
\BIBdecl

\bibitem{detection2017}
B.~Liang, G.~Fazekas, and M.~Sandler, ``Detection of piano pedaling techniques
  on the sustain pedal,'' \emph{Journal of the Audio Engineering Society},
  2017.

\bibitem{sustainPed}
------, ``Piano legato-pedal onset detection based on a sympathetic resonance
  measure,'' in \emph{Proc. EUSIPCO}, 2018.

\bibitem{Liang_2019}
------, ``Transfer learning for piano sustain-pedal detection,'' in \emph{Proc.
  IJCNN}, 2019.

\bibitem{TTtranscription}
Q.~Kong \emph{et~al.}, ``High-resolution piano transcription with pedals by
  regressing onsets and offsets times,'' \emph{arXiv preprint
  arXiv:2010.01815}, 2020.

\bibitem{hsiao2021compound}
W.-Y. Hsiao \emph{et~al.}, ``{Compound Word Transformer}: Learning to compose
  full-song music over dynamic directed hypergraphs,'' in \emph{Proc. AAAI},
  2021.

\bibitem{huang2018music}
C.-Z.~A. Huang \emph{et~al.}, ``{Music Transformer},'' in \emph{Proc. ICLR},
  2019.

\bibitem{emopia}
H.-T. Hung, J.~Ching, S.~Doh, N.~Kim, J.~Nam, and Y.-H. Yang, ``{EMOPIA}: A
  multi-modal pop piano dataset for emotion recognition and emotion-based music
  generation,'' in \emph{Proc. ISMIR}, 2021.

\bibitem{hawthorne2018enabling}
C.~Hawthorne \emph{et~al.}, ``Enabling factorized piano music modeling and
  generation with the {MAESTRO} dataset,'' in \emph{Proc. ICLR}, 2019.

\end{thebibliography}

\end{document}